\documentstyle[12pt]{article}
\begin{document}

\title{Anomalous behavior of the irreversible magnetization and time relaxation in YBa$_2$Cu$_3$O$_7$ single crystals with splayed tracks}
\author{A. Silhanek$^1$, D. Niebieskikwiat$^1$, L. Civale$^1$, M. A. Avila$^2$, O. Billoni$^3$ and D. Casa$^{1,*}$.\\
$^1$Comisi\'{o}n Nacional de Energ\'{\i}a At\'{o}mica-Centro At\'{o}mico\\
Bariloche and Instituto Balseiro, 8400 Bariloche, Argentina. \\
$^2$Inst. de Fisica Gleb Wataghin, UNICAMP, Campinas, Brazil\\
$^3$FAMAF, Univ. Nac. de C\'{o}rdoba, Argentina.}

\date{July 1, 1999}

\begin{abstract}
We have studied the angular dependence of the irreversible magnetization
and its time relaxation in YBa$_2$Cu$_3$O$_7$ single crystals with one or two families of columnar defects inclined with respect to the c-axis. At high magnetic fields, the magnetization shows the usual maximum centered at the mean tracks' orientation and an associated minimum in the normalized relaxation rate. In contrast, at low fields we observe an anomalous local minimum in the magnetization and a maximum in the relaxation rate. We present a model to explain this anomaly based on the slowing down of the creep processes arising from the increase of the vortex-vortex interactions as the applied field is tilted away from the mean tracks' direction. 

\end{abstract}

\section{Introduction}

Pinning of flux lines by columnar defects (CD) in high temperature superconductors has been of considerable
interest in the last years. It is well known that these correlated defects
yield a strong enhancement of flux trapping, in particular if the applied
field ${\bf H}$ is aligned with the tracks\cite{civale-91a,konczykowski-91,zhukov-97c,silhanek-98a}. When ${\bf H}$ is tilted away
from the linear defect direction beyond a lock-in angle ${\Theta_L}$,
vortices form staircase structures with kinks connecting
segments trapped in the columnar defects. The appearance of these kinks is expected to reduce the critical current $J_c$ and to produce a faster
relaxation. Thus, the angular dependence of the persistent current density $J$, should show\cite{nelson-92,blatter-94,hwa-93b} a peak at the CD's direction, as indeed observed in many cases\cite{civale-91a,zhukov-97c,silhanek-98a,kwok-92,grigorieva-92,li-93,klein-93,hardy-96,oussena-96,zhukov-97b,herbsommer-98}. 

However, Zhukov et al.\cite{zhukov-98} have recently shown that the angular dependence of the irreversible magnetization of YBa$_2$Cu$_3$O$_7$ (YBCO) crystals with CD along the c-axis exhibit a local minimum rather than a maximum for that field orientation. They found that this interesting and anti-intuitive behavior is related to geometrical effects; if the rotation axis is
parallel to the shortest side of a rectangular sample the minimum is observed, but if the axis coincides with the longest side the usual maximum in the angular dependence is recovered. Their interpretation also involves a sharp {\it increase} in the critical current density parallel to the rotation plane as the field is tilted away from the CD. 

Although the geometrical aspects of the anomalous behavior were convincingly demonstrated\cite{zhukov-98}, the origin of the increase of the current density as kinks proliferate is still very unclear. They speculate that it may be related to the appearance of helicoidal instabilities in the kink structure, but certainly other explanations cannot be discarded based solely on their results. One important fact to be taken into account is that the persistent currents determined in magnetization studies of HTSC is usually much smaller than $J_c$, as it is strongly reduced by thermal relaxation. Thus, the observed features are more likely to be related to differences in the activation energy of the excitations that dominate the depinning process for different angles.

In this paper we show that the anomalous dip is also visible in YBCO crystals with one or two families of aligned columnar defects {\it inclined} with respect to the c-axis. When only one family is present (all the defects are parallel) the local minimum is centered at the CD's direction. This result demonstrates that the anomaly is only due to vortex-tracks interactions, and the influence of crystal anisotropy or pinning by twin boundaries can be ignored. We also find that when two families of tracks are present (planar splay) only one minimum, centered at the mean defect's direction, is observed. Because in this case no kinks connecting pins of the same family are present in the angular range in between the two tracks' orientations, helicoidal instabilities are ruled out as a possible origin of the anomaly. To explore the nature of the thermal activation processes we performed time relaxation measurements as a function of angle. We show that the minimum in the irreversible magnetization is associated to a faster relaxation. We propose an alternative explanation of the anomalous angular dependence based on the reduction of the creep processes due to the increase of the vortex-vortex interactions as ${\bf H}$ is inclined with respect to the mean CD direction.

\section{Experimental}

We carried out magnetic studies of two YBCO single crystals grown from the
self flux method and oxygenated following the procedures described in ref.17. Both crystals were taken from the same batch, and display a $T_c=91.6K$ before irradiation. The two crystals have similar thickness $t \sim 15 \mu m$ and approximately rectangular shape, with dimensions $L \times s \sim 0.67 \times 0.22 mm^2$ for sample {\it A} and $L \times s \sim 0.58 \times 0.48 mm^2$ for sample {\it B}, where $L$ and $s$ are the long and short side respectively. 

Columnar defects were created by $309 Mev$ $Au^{+26}$ ion irradiation at 
the TANDAR accelerator in Buenos Aires, Argentina. In both cases the defects were introduced\cite{herbsommer-98} inclined with respect to the c-axis, with the irradiation plane (the plane formed by the c-axis and the irradiation direction) perpendicular to $s$. Sample {\it A} was irradiated at an angle $\Theta_D=10^{\circ }$ off the c-axis, and the
dose was equivalent to a matching field $B_{\Phi}=3T$. Sample {\it B} has two sets of columnar defects, one at $\Theta_{D1}=+5^{\circ }$ and the other at $\Theta_{D2}=+15^{\circ }$,
each one with a matching field $B_{\Phi 1}=B_{\Phi 2}=1.5T$. In this
way we obtain the same total dose ($3T$) and the same average angle for the
columnar defects ($10^{\circ }$) in both samples.

DC magnetization measurements were made in a Quantum Design SQUID magnetometer with a 5T magnet. The magnetometer is equipped with two sets of detectors, which allows us to record both the longitudinal ($M_l$) and the transverse ($M_t$) components of the
magnetization (parallel and perpendicular to ${\bf H}$ respectively). The
samples can be rotated {\it in situ} around an axis perpendicular to ${\bf H}$ using a home-made rotating holder\cite{casa}. 

To perform the magnetic measurements the crystals were carefully aligned with the rotating axis normal to the irradiation plane, in such a way that the condition ${\bf H}\parallel$ tracks could be achieved within $\sim 1^{\circ }$. This configuration also satisfies the geometrical condition (rotation axis parallel to the short side) required\cite{zhukov-98} to observe the minimum.

It is known that the measurement of $M_t$ in a QD magnetometer possesses some
difficulties arising from the presence of a spurious signal due to the
longitudinal component $M_l$ that is detected by the transverse pick-up
coils. This occurs when the sample is slightly off-center with respect to
the vertical axis of the coils, which is frequently the case. We have
completely and satisfactorily solved this problem. The solution includes an
initial alignment procedure and the external processing of the original
SQUID output signal using software developed ad-hoc. All the details
related with the hardware and software of the sample rotation system will be
presented elsewhere\cite{casa}.

We performed isothermal magnetization loops maintaining a fixed value
of the angle ${\Theta}$ between the normal to the crystal (c-axis) and the applied
field direction, and recording both components $M_l(H)$ and $M_t(H)$. We use the widths of the hysteresis $\Delta M_l(H)$ and $\Delta M_t(H)$ to calculate the modulus $M_{i}=\frac{1}{2} \sqrt{\Delta M_l^2+\Delta M_t^2}$ and direction of the irreversible magnetization vector $\bf {M_i}$. Loops were recorded up to $H=5T$ in all cases. As $H$ is reduced from this maximum field, the non-equilibrium currents that generate the critical state profile start to reverse direction\cite{campbell-72}. The formation of a fully developed critical state of the oposite sign occurs after a field decrease of the order of $\Delta H \sim 2H^* \sim Jt$, where $H^*$ is the self field\cite{daeumling-89}. This situation is clearly identified as $M_l(H)$ and $M_t(H)$ reach the field-decreasing branch of the loop. We have carefully checked that all the $\Delta M$ data shown in this work correspond to the difference between two oposite fully developed critical states, thus $M_i$ can be easily related to the persistent currents. After each loop is finished the sample was rotated, warmed up above $T_c$ and then cooled down to the working temperature in zero field. In this way, the initial Meissner response was recorded for each angle. 

As the non-equilibrium currents in thin samples are strongly constrained to flow parallel to the sample surface, $\bf {M_i}$ points almost perpendicular to the surface \cite{silhanek-98a,hellman-92,zhukov-97a,candia-98} in a wide angular range of applied field $0 < \Theta < \Theta_c$. For both crystals the critical angle $\Theta_c \sim \arctan(L/t) > 87^{\circ}$, and we indeed confirmed that $\bf {M_i}$ was normal to the sample surface within our $1^{\circ}$ resolution for all the angles shown in the present work. (The angle $\Theta$ was determined independently using the Meissner slopes, as described in ref. 21).

\section{Results and Discussion}

Figure 1 shows the angular dependence of the modulus of the irreversible magnetization as a
function of ${\Theta}$ for crystal {\it A} at two temperatures. The main feature of this figure is the evident asymmetry with respect to the
c-axis, which is due to the uniaxial vortex pinning produced by the inclined columnar defects. The anomalous {\it minimum} is apparent at both temperatures. This dip is centered at the tracks' direction $\Theta_D=10^{\circ}$ (except at very low fields, as discussed below).
At $T=35K$ the minimum is visible for all values of the applied field. Its depth first increases with $H$, reaches a maximum at 
$H \sim 1T$ and then progressively decreases. At $T=70K$, on the other hand,  the dip is only observed at 
low fields, its depth monotonically decreasing with $H$ until the behavior switches to the well-known peak at higher fields. At this temperature it becomes clear that the dip is "mounted" over the broader usual maximum centered at the tracks' direction. The angular width of the minimum decreases with both temperature and field increase.

Also visible in figure 1 is a shift of the dip from the tracks´ direction towards the c-axis. This shift occurs for the lowest fields at both temperatures, although it is not shown at $70K$ for clarity. In a previous work we have shown that this effect is related to the misalignment between the internal flux density $\bf{B}$ (which represents the vortex direction) and $\bf{H}$, due to the anisotropy\cite{silhanek-98a}. From now on we will concentrate on the field regime where $\bf{B} \parallel \bf{H}$.

As a first step to investigate the origin of the anomalous minimum, we must determine the geometrical relation between $M_{i}$ and the non-equilibrium currents flowing in our samples, as a function of $\Theta$. To that end we will use the extended Bean critical state model for in-plane anisotropic currents\cite{gyorgy-89}. For our thin and approximately rectangular crystals we obtain 

\begin{equation}
M_i=\frac{J_1 s}{20} (1-\frac{s}{3L} \frac{J_1}{J_2}) 
\end{equation}
where $J_1(\Theta)$ and $J_2(\Theta)$ are the current densities (constrained to flow in the plane of the sample) parallel to the long and short sides of the sample respectively, as sketched in figure 2(a). Eq. 1 is valid provided that $J_1/J_2<L/s$.

When $\bf{H}$ is parallel to the c-axis all the currents are perpendicular to $\bf{H}$, thus the Lorentz force on the vortices is maximum for both current directions and then
$J_1(\Theta=0)$ and $J_2(\Theta=0)$ are equal to the critical currents $J_{c1}$ and $J_{c2}$ respectively. Note that, in contrast to the case analyzed by Zhukov et al., $J_{c1}(\Theta=0)$ and $J_{c2}(\Theta=0)$ in our case are different, due to the inclination of the tracks with respect to the c-axis\cite{schuster94a}. 

We must now analyze how $M_i$ is expected to behave when $\bf{H}$ is tilted from the c-axis. In this case $J_2(\Theta)$ remains perpendicular to $\bf{H}$ and thus $J_2(\Theta)=J_{c2}(\Theta)$. On the other hand, only the component of $J_1(\Theta)$ perpendicular to $\bf{H}$ contributes to the Lorentz force\cite{brandt-94b}, and thus $J_1(\Theta)=J_{c1}(\Theta)/cos(\Theta)$. Thus, if $J_{c1}$ and $J_{c2}$ were independent of $\Theta$, the $M_i(\Theta)$ would increase as $\Theta$ grows from $0$ to $90^{\circ}$. However, the minimum originated in this effect is centered at the c-axis and not at the direction of the columnar defects. Moreover, the observed minimum in $M_i$ is much sharper than $1/cos(\Theta)$. 

We then conclude that, also in our inclined defects' configuration, any explanation of the minimum must involve an increase of either $J_{c1}(\Theta)$ or $J_{c2}(\Theta)$ as $\bf{H}$ deviates from the tracks' direction. As the second possibility has been ruled out by the results obtained when the crystals are rotated around the longer axis\cite{zhukov-98}, we will focus our analysis in the possible reasons for the sharp increase of $J_{c1}(\Theta)$.

The interpretation suggested by Zhukov et al. is based on the appearance of kinks connecting nearby tracks when the field is inclined with respect to them. When the sample is rotated around its shorter axis, those kinks are on the average perpendicular to $J_2$ and have a component parallel to $J_1$. It was speculated that in the force free configuration associated to $J_1$ kinks may develop helicoidal instabilities\cite{indenbom-97}, thus resulting in an increase of $J_1$. 

To check this possibility, we repeated the study on crystal {\it B}. The key difference in this case, as we will demonstrate below, is that for any field orientation in between the two families of tracks, kinks connecting tracks of the same orientation do not exist, and consequently {\it  helicoidal instabilities cannot develop}. 

Figure 3 shows $M_i (\Theta)$ at $T=60K$ for crystal {\it B}. The anomalous minimum at low fields is also clearly visible in this case, switching to the usual maximum at high fields. Now both the minima and the maxima are centered at the mean tracks' orientation, $\Theta=10 ^{\circ}$. These $M_i (\Theta)$ curves cannot be satisfactorily adjusted by superposition of two minima (or maxima) centered at $\Theta_{D1}$ and $\Theta_{D2}$, indicating that the observed behavior results from the combined interaction of vortices with both families of tracks\cite{herbsommer-98}.

We now analyze the vortex structure in this crystal as a function of $\Theta$. For $5^{\circ}<\Theta<15^{\circ}$ vortices may zigzag between tracks of different families. If two such tracks physically intersect, no kink is required to connect the two pinned vortex segments. If, on the contrary, the two tracks are not in the same plane, a kink connecting both pinned segments must exist. Let us consider a track of the family $\Theta_{D2}=15^{\circ}$. The number of tracks of the other family ($\Theta_{D1}=5^{\circ}$) that approach to it within a distance $D$ of its axis is equal to the number of such tracks that cross the rectangle of area
$A=2Dt[\tan(15^{\circ})-\tan(5^{\circ})] \sim 0.34 t D$, as seen in figure 4(a). 

We can estimate such number as $n=A B_{\Phi 2}/\Phi_0$. This gives, for instance, an average of 23 "close approaches" within a distance $D=6 nm$, which approximately corresponds to the diameter of the tracks. The average distance between such crosses is about $\delta \sim t/n \sim 700 nm$. The same estimate for $\delta$ can be obtained using the more elaborate analysis of Hebert et al.\cite{hebert-98}. 

The above analysis indicates that for field orientations $\Theta_{D1}<\Theta<\Theta_{D2}$ and at low fields, when vortex-vortex interactions are small, the energetically most convenient configuration for most of the vortices is to zigzag among tracks that intersect within their diameter ($D \leq 6 nm$), thus {\it  not forming kinks}. Of course, considering the random distribution of tracks there is a probability that some kinks will exist, but the number of them will be negligible as compared to the case of one single family of tracks at similar field and inclination.

The situation in this angular region changes as $H$ increases, because the vortex-vortex interactions tend to inhibit the transverse displacements required to zigzag. For instance, for a field direction $\Theta=10 ^{\circ}$, pinned segments of length $\delta$ imply transverse displacements of $\sim \delta sin(5^{\circ}) \sim 60 nm$. When the distance between vortices decreases to around this value (that corresponds to $H \sim 0.5 T$), it becomes energetically convenient to form some kinks connecting tracks of different families that do not intersect (i.e., separated by $D > 6 nm$). This reduces $\delta \propto 1/D$, and consequently the transverse excursions. 
However, as long as $D$ remains smaller than the average distance between tracks of the same family, $d \sim 36 nm$ in our sample, it will still be convenient to form kinks between tracks of different families (inter-family kinks) rather than within the same family (intra-family kinks). 

These inter-family kinks cannot develop helicoidal instabilities, for several reasons. First, their lengths are in the range of $\sim 10 nm$, too short to entangle. Second, according to Indenbom et al.\cite{indenbom-97} these instabilities are only visible in {\it extremely low pinning crystals}, which is certainly not our case. Finally, the kinks have their main component perpendicular to the plane of irradiation, which in our geometry means parallel to the axis of rotation. Thus, the force free configuration required for the appearance of the instabilities may only be produced by $J_2$, instead of $J_1$, contrary to the original argument. Consequently, the observation of the minimum in this angular range, clearly seen in figure 3, rules out the possibility that it is associated with the helicoidal instabilities.

For $\Theta < \Theta_{D1}$ and $\Theta > \Theta_{D2}$ the nature of the vortex structure changes. It will now be energetically favorable for vortices to form staircases with segments pinned mainly in tracks of one family, connected by kinks of the same type and orientation as those formed in samples with all pins in a single direction. Therefore, to a first approximation we can ignore the second family of tracks, and we have a situation similar to that of sample {\it A}. However, it is clear from fig. 3 that no hint of a change in behavior is seen either at $\Theta = 5^{\circ}$ or $\Theta = 15^{\circ}$. Thus, the minimum appears to be independent of the presence or absence of intra-family kinks. 

Once the helicoidal instabilities have been discarded as possible sources of the anomalous minimum, we must search for an alternative explanation. A fact that we have not considered up to now is that, due to the large influence of thermal fluctuations on the vortex dynamics in HTSC, the persistent current density $J$ determined through dc magnetization measurements in the typical time scale of SQUID magnetometers is much smaller than the "true" critical current density $J_c$. This suggests that the anomalous minimum may be related to the angular dependence of the time relaxation of $J$.

To confirm this possibility we have measured the normalized time relaxation rate of the irreversible magnetization, $S=-dLn(M_i)/dLn(t)$, for the splayed sample {\it B} as a function of $\Theta$. Measurements were performed at $T=60K$ for two values of field: $H=3T$, where $M_i (\Theta)$ shows the usual maximum at the mean tracks' direction, and $H=0.5T$, where $M_i (\Theta)$ exhibits the anomalous minimum. The curves $S(\Theta)$ are presented in figure 5(a) and 5(b) respectively, together with the corresponding $M_i (\Theta)$ data, already shown in fig. 3.

Before we discuss the data shown in fig. 5, we must analyze how the quantity $S$ defined above relates to the normalized relaxation rates of both current densities flowing through the crystal, $S_1=-dLn(J_1)/dLn(t)$ and $S_2=-dLn(J_2)/dLn(t)$. Operating on eq. (1), we obtain

\begin{equation}
S(\Theta)=(1-K(\Theta))S_1 (\Theta)+K(\Theta)S_2 (\Theta)
\end{equation}

Thus, the influence of the relaxation rate of $J_1$ and $J_2$ on the global rate is weighted by the angle dependent geometrical factor

\begin{equation}
K(\Theta)=\frac {s J_1}{3L J_2 - s J_1}
\end{equation}

The range of validity of Eq. 2 is the same as eq. 1, i.e. $s J_1 < L J_2$, or equivalently $K<1/2$. The condition $K=1/2$ corresponds to the change in the shape of the inverted roof, from that of fig. 2(a) to that of fig. 2(b), that has been identified\cite{zhukov-98} with the maximum in $M_i (\Theta)$. 
On the other hand, $K$ is minimum at $\Theta=10^{\circ}$, where $J_1/J_2$ reaches its smallest value. If we estimate that $J_1(\Theta=10^{\circ})\sim J_2(\Theta=10^{\circ})$, then $K(\Theta=10^{\circ}) \sim 1/8$. Thus, in the angular range of the minimum, $S$ mostly reflects the behavior of $S_1$.

The high field behavior shown in figure 5(a) is in agreement with the theoretical expectations\cite{nelson-92}: 
$S$ decreases as the field orientation approaches the tracks' direction, due to the growth of the pinned fraction of the vortices and consequent increase of the activation energy. The observation of a single minimum at the mean direction of the tracks is well described by the scenario discussed by Hwa et al.\cite{hwa-93a}, according to which the forced entanglement of the vortices in the angular range $5^{\circ}<\Theta<15^{\circ}$ tends to inhibit the thermal relaxation. 

Figure 5(b) shows a different behavior. Here we observe a narrow and small peak mounted over a larger minimum, both centered at $10^{\circ}$. The main minimum (which is wider than at $H=3T$) corresponds once again to the increase of the pinned fraction. The central peak, on the other hand, is a new manifestation of the anomalous behavior. As we showed above, in this angular region $S$ is basically a measure of $S_1$. We conclude that the anomalous increase of $J_1$ as $\bf {H}$ is tilted away from the mean tracks' direction is a consequence of the reduction of $S_1$. It is important to note that the normalized relaxation rate is a very fundamental parameter of the vortex dynamics that characterizes the pinning and creep regimes, and is rather insensitive to the pinning details\cite{blatter-94}. On the contrary, the persistent current density is a more derived variable that depends on the time scale of the measurement. Thus, the basic concept is that the minimum in $J_1(\Theta)$ is a result of the maximum in $S_1(\Theta)$, and not the other way around. The goal now is to find the reason for this unexpected behavior of $S_1$.

It is well known that the increase of vortex-vortex interactions usually results in a decrease of the normalized relaxation rate, which manifests in a larger glassy exponent $\mu$ of the collective creep regime as compared to the single vortex creep. This stiffening of the vortex matter due to elastic interactions is very general, and rather independent of the pinning details, so it occurs both for correlated and random disorder. This suggests that the observed decrease in $S_1$ as we tilt the field away from the mean tracks´ direction may be due to the increase of the interactions. Strong support for this interpretation arises from two distinctive features of the $M_i (\Theta)$ data. 

First, we note in fig. 6(a) that, right at the mean tracks´ direction where the minimum occurs, $M_i$ at low fields is independent of $H$. This is clearly seen in the inset of fig. 6(a), where $M_i (8.4^{\circ})$ for crystal {\it B} at $T=60K$ is plotted as a function of $H$. The field independent $M_i$ regime, which is characteristic of a system of non-interacting vortices, extends up to 
$H \sim 1T$, i.e., it roughly coincides with the field range where the minimum occurs.

In the second place, if we tilt the field away from the tracks we observe that the increase of $M_i (\Theta)$ is steeper the higher $H$ is. This means that in the proximity of the minimum $M(H)$ (at fixed $\Theta$) grows with $H$, i.e. $M(H)$ exhibits a fishtail shape. Fishtail loops (observed in many HTSC compounds) have been attributed to a variety of origins. In some cases\cite{civale-94} the increase of $M$ with $H$ has been shown to originate in the reduction of the relaxation rate with increasing $H$, which is a consequence of the increase of the vortex-vortex interactions. The increase of $M_i$ with both the tilt angle $\Delta \Theta = \Theta - 10^{\circ}$ and $H$ suggests {\it a common origin of both dependencies}. This becomes apparent in fig. 6(b), where the $M_i (\Theta ,H)$ data of fig. 3 is replotted as a function of the field component perpendicular to the mean tracks´ direction, $H_{\perp}=H sin(\Delta \Theta)$. In the field range $H \leq 1T$ we observe that the various curves have the same curvature around the minimum.

In summary, the scenario that emerges from the angle dependence of $S_1$ and the angle and field dependence of $M_i$ is the following. In the field range of the anomalous minimum, and for $H$ parallel to the mean tracks´ direction, vortex-vortex interactions are small. Those interactions increase with $H_{\perp}$, thus resulting in a reduction of $S_1$ and the consequent increase of $M_i$ measured at fixed time.

The same features in $M_i(\Theta)$ are observed at $T=70K$ in crystal A: 
$M_i(\Theta = 10^{\circ})$ is field independent up to $\sim 1T$, and the curves $M_i(\Theta)$ at different fields have the same curvature around the minimum when plotted as $M_i(H_{\perp})$. 

The reason for the increase of the interactions with $\Theta$ at constant $H$, i.e., at constant average distance between vortices, is not obvious. In the case of zigzagging or staircase vortices, the distance between neighbors varies along the field direction. As the vortex-vortex repulsion is a highly nonlinear function of their separation, the strength of the interactions depends not only on their average distance but also on the amplitude of the transverse displacements. A detailed analysis of the interaction energy as a function of $\Theta$ thus requires the complete computation of all the 3D configurations involved. This is a very difficult problem, that we will not attempt to solve here. However, we will now present a simple estimate that shows how the interactions in the presence of splay defects and at low fields may increase as $H$ is tilted away from the mean tracks' direction.

As discussed above, for $5^{\circ}<\Theta<15^{\circ}$ and low fields, it is useful to consider that vortices zigzag within a {\it planar grid} of tracks as shown in fig. 4(b). The maximum displacement of the vortex perpendicular to the field direction, $R$, as a function of $\Theta$, can be estimated as $R \sim \delta sin(5^{\circ}+\mid \Delta \Theta \mid)$. 
This relation shows that $R (\Theta)$ is minimum at the mean tracks' direction $\Theta=10^{\circ}$. As a result, the {\it minimum distance} between two adjacent vortices decreases with $\Delta \Theta$, even when the {\it average distance} between them remains constant, thus producing an increase of the interaction.

For sample {\it A}, with only one family of tracks, the transverse displacements must be calculated differently. Now the zigzag-vortices, without kinks, must be replaced by staircase-vortices with kinks connecting parallel tracks. The orientation of a kink (the angle between the kink and the tracks) depends on the pinning energy of the two adjacent tracks, thus the dispersion in the pinning energy of the columnar defects results in a dispersion of the kinks´ orientations\cite{silhanek-98a}. The larger the pinning energy, the closer to the ab-plane is the kink. In a previous study we have shown that, when $\Theta$ exceeds the angle of a particular kink, such kink disappears and the vortex involved becomes trapped by stronger tracks connected by a longer kink, closer to the ab-plane. This process generates a progressive increase of both the average kink length and the deviation from $\Theta_D$ as $\Delta \Theta$ grows, which again results in a decrease of the minimum distance between adjacent vortices, at constant $H$.

We can use the analysis presented in the two previous paragraphs for the crystals {\it B} and {\it A}, with and without splay respectively, to compare the situation for $\Theta=10^{\circ}$ in both cases. At this angle, the average transverse displacements of the zigzagging vortices of crystal {\it B} is much larger than in crystal {\it A}, where vortices are expected to be locked. Thus, the interactions at the same $H$ should be larger for crystal {\it B}. This is consistent with the observed values of the field required to switch from the anomalous minimum to the maximum in $M_i (\Theta)$. Although such field for each sample decreases with temperature, 
it is higher in sample A at 70 K ($\sim 3.5T$, see fig. 1(b)) than in sample B at 60 K ($\sim 1.5T$, see fig. 3).

\section{Conclusion}

The similarities in the behavior of the crystals with parallel and splayed tracks indicate that the physics involved in the anomalous minimum is rather independent of the details of the vortex configurations. The local maximum in the angular dependence of the normalized relaxation rate demonstrates that the minimum in $M_i(\Theta)$ is due to a stiffening of the vortex matter as $\bf H$ is tilted away from the mean tracks direction. We attribute such effect to the increase of the vortex-vortex interactions arising from the enlargement of the transverse vortex displacements, but clearly further studies are required for a complete understanding of the phenomenon.

\section{acknowledgment}

Work partially supported by ANPCyT, Argentina, PICT 97 No.01120, and CONICET PIP 4207. A.S. is member of CONICET. We acknowledge useful discussions with J. Luzuriaga and H. Pastoriza. M.A.A. and O.B. thanks Abdus Salam ICTP for supporting their participation in the Experimental Workshop of High Temperature Superconductors and Related Materials, Bariloche, Argentina, September 1998.

\section{References}

\bibliographystyle{prsty}
$^*$Present Address: Department of Physics, Princeton University, New Jersey 08544, USA.

\noindent
Figure 1: Irreversible magnetization $M_i$ for crystal {\it A} (with a single family of tracks), as a function of the applied field angle, $\Theta$, at several fields and at temperatures (a) T=35K and (b) T=70K.

\noindent
Figure 2: Sketches of the critical state profiles for different angular regimes. The pictures (a) and (b) shows the shape of roof pattern for $\frac {J_1}{J_2} < \frac {L}{s}$ and $\frac {J_1}{J_2} > \frac {L}{s}$, respectively. The rotation axis is parallel to the shortest side of the sample, and perpendicular to the irradiation plane.

\noindent
Figure 3: Angular dependence of the irreversible magnetization $M_{i}(H)$ at several fields for sample {\it B} (with splayed defects), at T=60K.

\noindent
Figure 4: (a) Sketch used to estimate the number of defects of family 1 that approximate to one track of family 2 within a distance $D$, in a crystal of thickness $t$. (b) Transverse excursions $R$ of a vortex in a planar grid of splayed columnar defects, as a function of angle.

\noindent
Figure 5: Angular dependence of the normalized relaxation rate $S$ (open symbols) and irreversible magnetization $M_{i}(H)$ (solid symbols) for sample {\it B} (with splayed defects), at T=60K and fields (a) $H=30kG$ and (b) $H=5kG$.

\noindent
Figure 6: (a) Blow up of the data showed in fig.3 in the region of the minimum at low fields. (b) Irreversible magnetization $M_{i}$ as a function of the field component normal to the tracks' direction, $H_{\perp}$. The inset shows $M_{i}$ as a function of the applied field $H$ at the mean tracks direction.

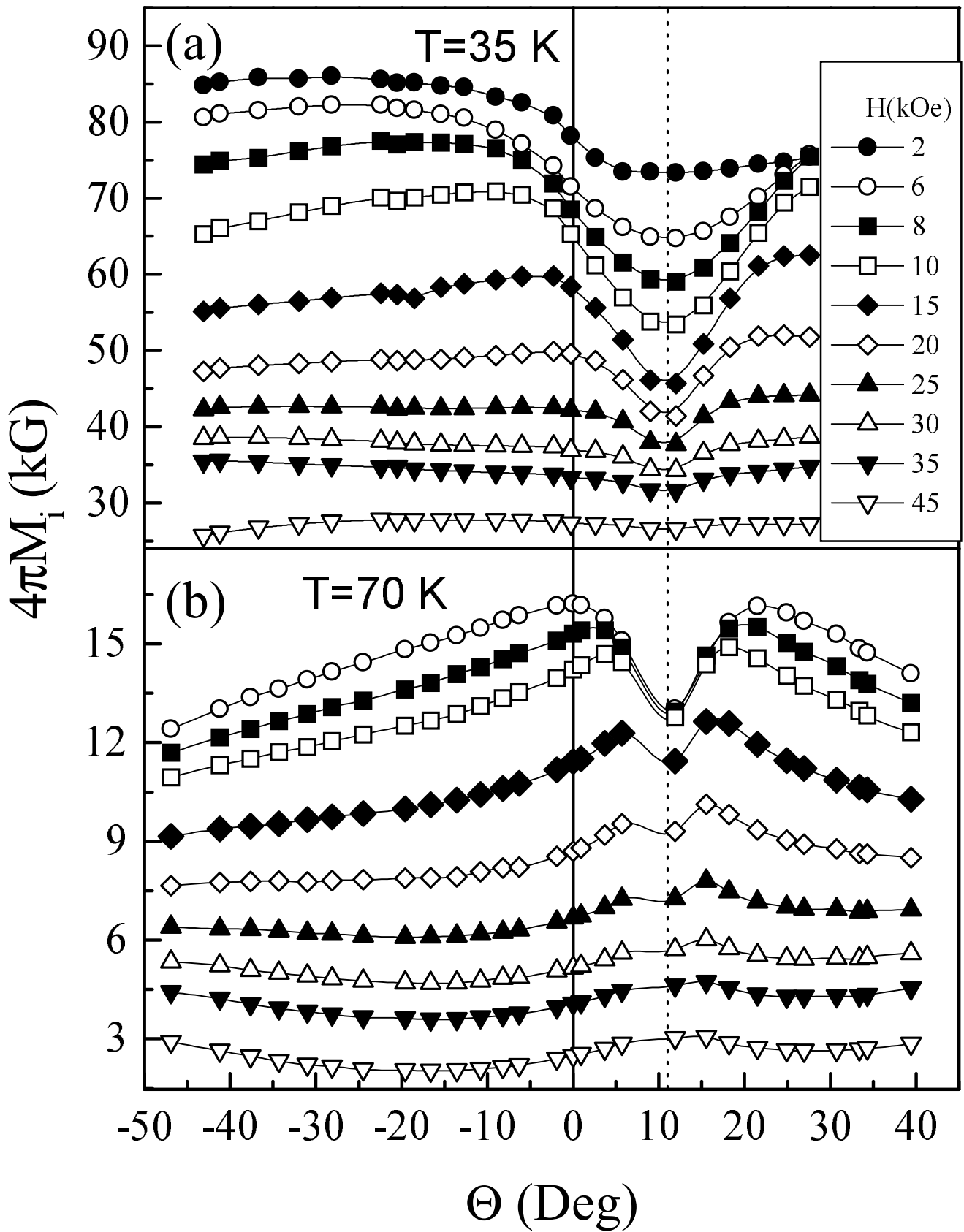
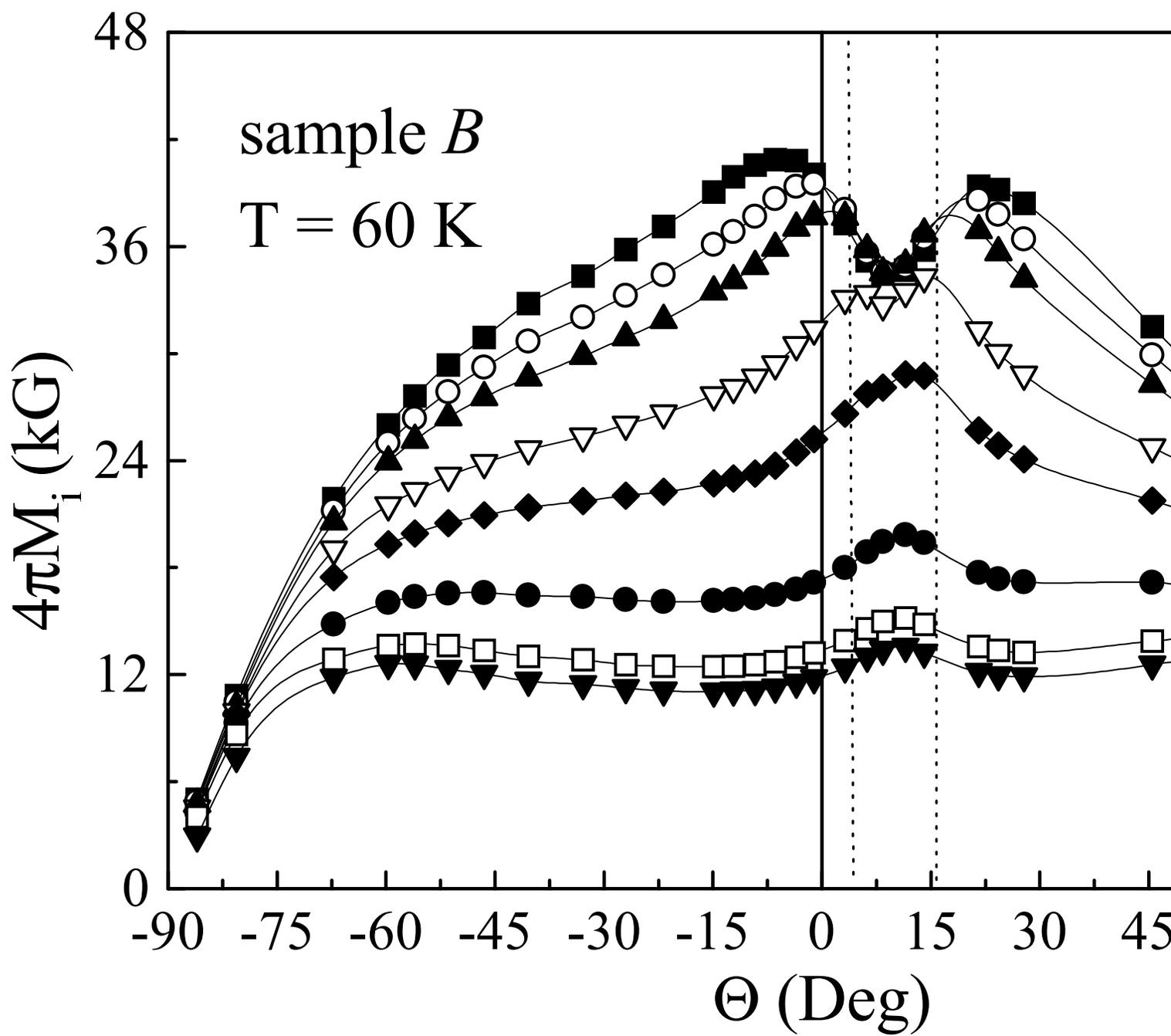
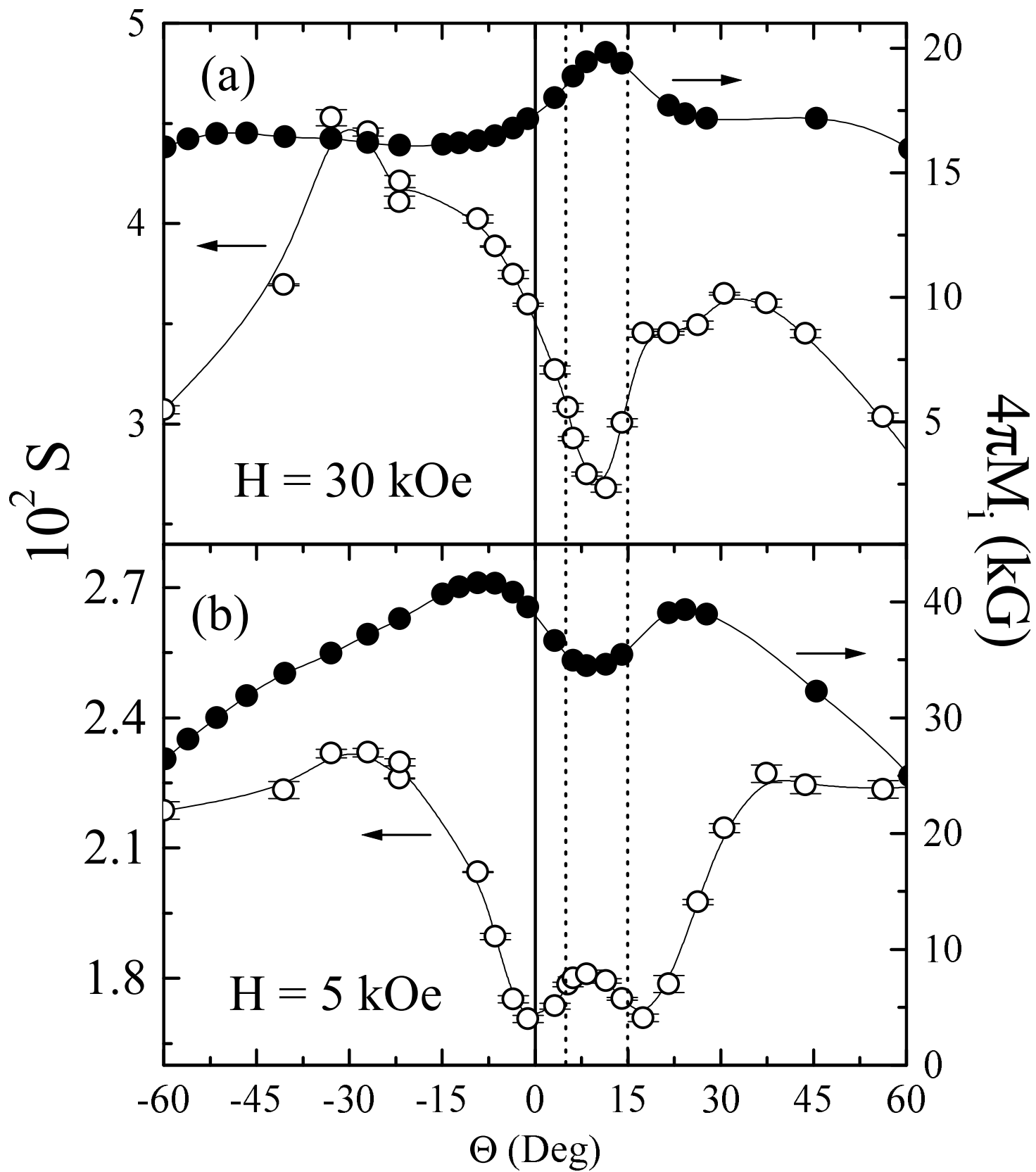
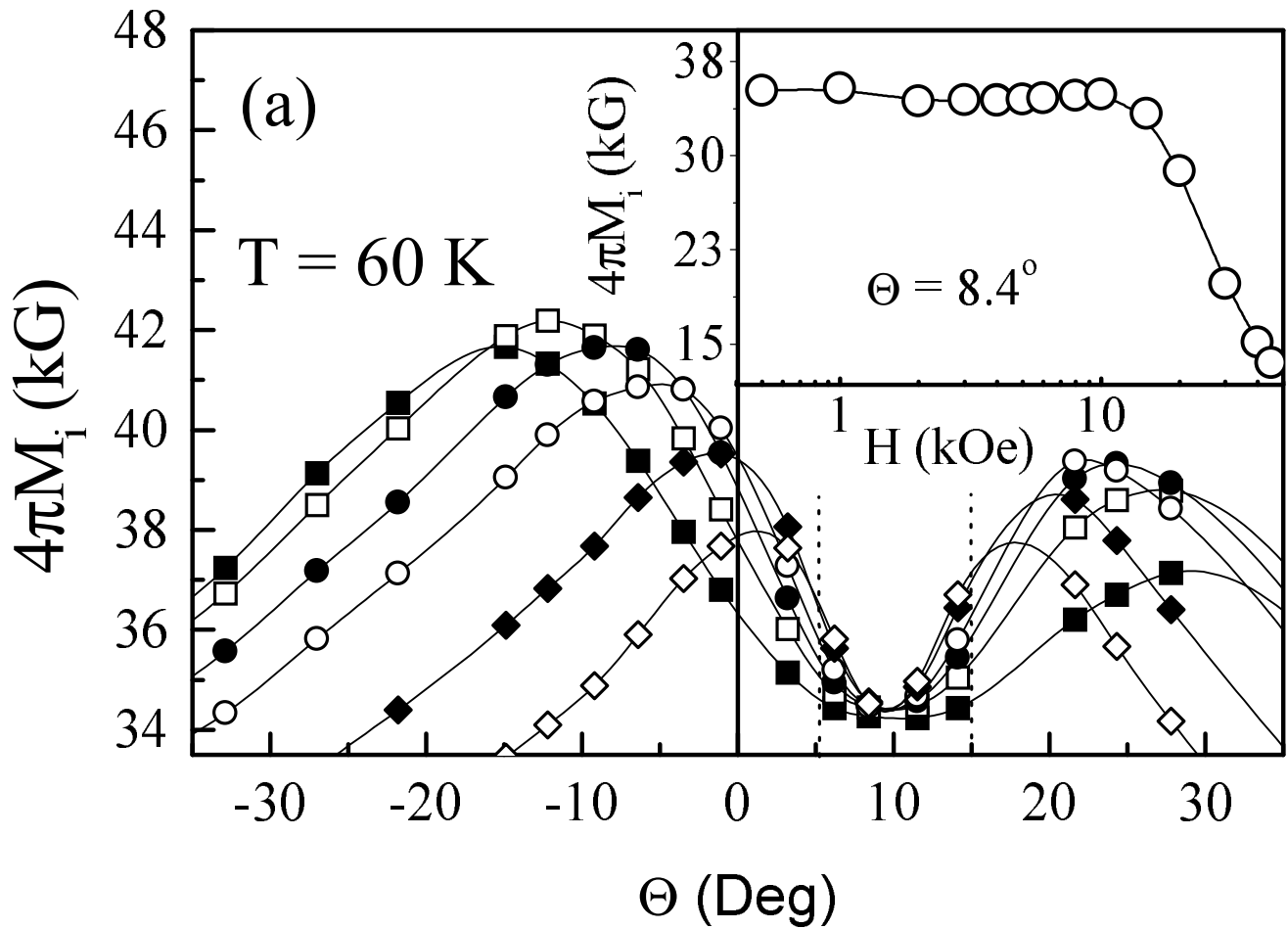
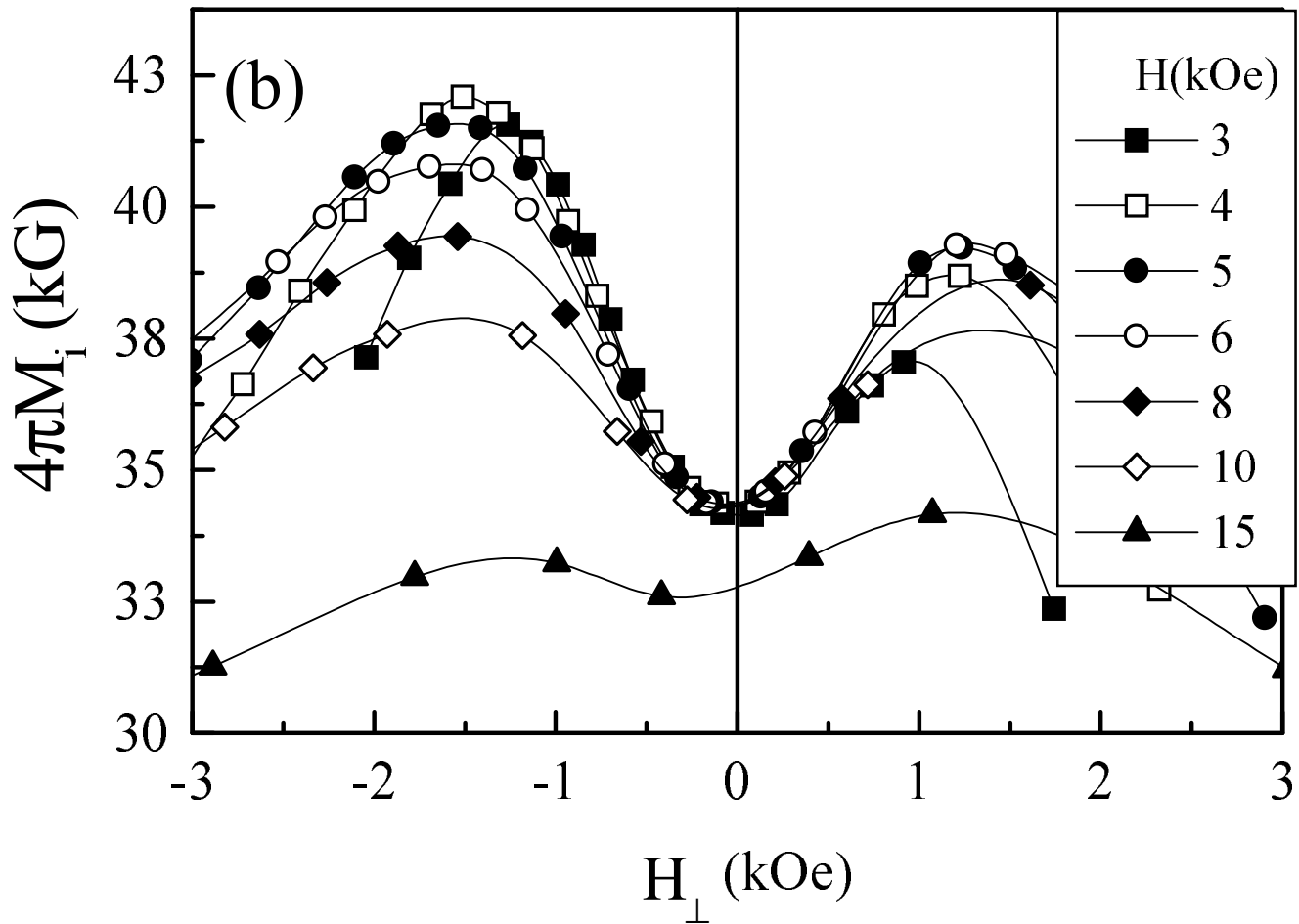
\end{document}